\documentclass[fleqn,10pt]{wlscirep}
\usepackage[utf8]{inputenc}
\usepackage[T1]{fontenc}
\usepackage{caption}
\title{Table-top interferometry\\ on extreme time and wavelength scales}

\author[1]{S. Skruszewicz}
\author[1]{A. Przystawik}
\author[1]{D. Schwickert}
\author[1]{M. Sumfleth}
\author[1]{M. Namboodiri}
\author[2]{V. Hilbert}
\author[2,3]{R. Klas}
\author[4]{P. Gierschke}
\author[2]{V. Schuster}
\author[5]{A. Vorobiov}
\author[5]{C. Haunhorst}
\author[5]{D. Kip}
\author[2,3,4]{J. Limpert}
\author[2,3,4]{J. Rothhardt}
\author[1,6,*]{T. Laarmann}
\affil[1]{Deutsches Elektronen-Synchrotron DESY, Notkestraße 85, Hamburg, 22607, Germany}
\affil[2]{Institute of Applied Physics, Friedrich-Schiller-University Jena, Albert-Einstein-Straße 15, 07745 Jena, Germany}
\affil[3]{Helmholtz Institute Jena, Fröbelstieg 3, 07743 Jena, Germany}
\affil[4]{Fraunhofer Institute for Applied Optics and Precision Engineering, Albert-Einstein-Straße 7, 07745 Jena. Germany}
\affil[5]{Faculty of Electrical Engineering, Helmut Schmidt University, Holstenhofweg 85, Hamburg 22043, Germany.}
\affil[6]{The Hamburg Centre for Ultrafast Imaging CUI, Luruper Chaussee 149, Hamburg 22761, Germany.}
\affil[*]{tim.laarmann@desy.de}
\begin{abstract}
Short-pulse metrology and dynamic studies in the extreme ultraviolet (XUV) spectral range greatly benefit from interferometric measurements. In this contribution a Michelson-type all-reflective split-and-delay autocorrelator operating in a quasi amplitude splitting mode is presented. The autocorrelator works under a grazing incidence angle in a broad spectral range $(\mathrm{10\,nm - 1\,\mu m})$ providing collinear propagation of both pulse replicas and thus a constant phase difference across the beam proﬁle. The compact instrument allows for XUV pulse autocorrelation measurements in the time domain with a single-digit attosecond precision and a useful scan length of about 1\,ps enabling a decent resolution of $\mathrm{E/\Delta E}=2000$. Its performance for selected spectroscopic applications requiring moderate resolution at short wavelengths is demonstrated by characterizing a sharp electronic transition at $\mathrm{26.6\,eV}$ in Ar gas. The absorption of the $\mathrm{11^{th}}$ harmonic of a frequency-doubled Yb-fiber laser leads to the well-known $\mathrm{3s3p^{6}4p^{1}P^{1}}$ Fano resonance of Ar atoms. We benchmark our time-domain interferometry results with a high-resolution XUV grating spectrometer and find an excellent agreement. The common-path interferometer opens up new opportunities for short-wavelength femtosecond and attosecond pulse metrology and dynamic studies on extreme time scales in various research fields. 
\end{abstract} 
\begin{document}

\flushbottom
\maketitle
%
%
\thispagestyle{empty}

\section{Introduction}

Interferometry in the extreme ultraviolet spectral range (XUV) plays a key role in attosecond pulse metrology for more than 20 years \cite{Bellini1998}. It has led to a wide range of applications ranging from attosecond spectroscopy \cite{Smirnova2009} to coherent diffractive imaging on the nanoscale using table-top sources \cite{Meng2015,Rothhardt2018}. It is the capability of interferometric methods to access the relative phases and amplitudes of interfering waves known from optics that is used to track propagating quantum objects, such as charge or spin waves in matter on extremely short distances and corresponding ultrafast time scales \cite{Gonzalez2020}. Along these lines, phase-sensitive light wave metrology in time and frequency domain goes hand-in-hand with advanced coherent spectroscopy and imaging applications using electrons, ions or photons as observable. Experimental breakthroughs, such as watching real-time electronic wave packet dynamics in atoms \cite{Wituschek2020}, orbital tomography in molecules \cite{Bertrand2013} and optical coherence tomography of solids \cite{Fuchs2017}, rely on controlling or detecting the phase of an electromagnetic wave with high precision \cite{Azoury2021,Uzan2020}.\\ 

In typical spectroscopic studies, the signal intensity is recorded, while the phase information is lost. High-resolution XUV spectrometers provide a resolution ($\mathrm{E/\Delta E}$), which depends on the diffraction grating period, i. e. the number of illuminated groves and on the wavelength of the light. Typically, it is on the order of $\mathrm{E/\Delta E=10^{\,3}}$ \cite{Wunsche2019}, while special custom-made gratings can reach a resolution of up to a few times $10^{\,4}$. Although these optical elements are not off-the-shelf and very expensive, they are now utilized in large scale soft x-ray synchrotron beamlines especially for spectroscopic techniques demanding high resolution, such as resonant inelastic x-ray scattering (RIXS) (see for instance \cite{Singh2021}). An alternative approach is Fourier-transform (FT) spectroscopy. By its nature, it enables characterization of radiation in both time and frequency domain using a single instrument. The spectral resolution of such a system depends on the maximum optical-path difference between replicas of the beam propagating in both interferometer arms \cite{Bates1976}. Thus, it can exceed the spectral resolution of the grating-based analyzers by orders of magnitude, if very large delays can be used \cite{Oliveira2011}. Moreover, the resolution is uniform across the entire photon energy range, which usually is not the case for grating-based spectrometers.\\

The standard geometry of FT spectrometers employs the typical Michelson-interferometer beam path. Its essential component is a beam splitter, which permits for splitting the field amplitude of the incoming laser pulse in two identical parts propagating along two perpendicular interferometer arms. Combining the two replicas with parallel wavefronts on the detector maximizes the interference contrast. The transfer of the two-arm concept into the short-wavelength range has been demonstrated for specific wavelengths based on multi-layer optics \cite{Hilbert2013,Hilbert2014}. The main challenge for performing FT spectroscopy in the XUV and beyond is, however, the design of a beam splitter covering a broad spectral range, reaching high transmission and maintaining a plane wavefront. To meet these challenges, alternative schemes have been developed in recent years based on either engineering two source points in close proximity with variable pulse delay or special split-and-delay units, applied to a single radiation source. In the former scheme, two mutually coherent XUV pulses separated in space and time can be generated by harmonic generation from two laser pulses focused into a gas jet with the same separation \cite{Kovacev2005,Tzallas2003,Jansen2016}. The latter concept is based on a conventional half-mirror split-and-delay unit \cite{Mashiko2003,Mashiko2020}. Both approaches have already been employed to characterize attosecond pulse trains in time and frequency domain \cite{Chen2014,Jansen2016} and for spectroscopy at synchrotron radiation facilities \cite{Oliveira2011}. The major drawback of these concepts is that the two XUV pulses do not propagate collinearly. Due to the partial overlap in the far field, only a small fraction of the lateral beam profile can be used to evaluate the interference signal. Because the pulses interfere at small skew angles, the phase between beams changes continuously along the intersection path in the focal volume. Consequently, the interference contrast in these non-collinear geometries is typically much weaker than in Michelson-type interferometers.\\

In this contribution we demonstrate FT spectroscopy in the XUV spectral range by employing a recently developed Michelson-type all-reflective split-and-delay autocorrelator design \cite{Gebert2014}. The instrument is based on two interleaved lamellar mirrors, which generate two replicas of XUV pulses propagating collinearly. It enables quasi amplitude division, resolves interferometric autocorrelation signals and can be applied in modern Ramsey-like experimental schemes \cite{Eramo2011}. It implies that the device can be used to acquire phase information in some cases, for instance, if implemented in nonlinear detection schemes that go beyond measuring field-autocorrelation traces in spectroscopic applications. Of course, the latter are better off at continuous or pseudo-continuous sources like synchrotrons \cite{Oliveira2011}. The common-path laser interferometer allows for tight focusing of phase-coherent pulse replicas while keeping the fringe contrast at maximum in the focal volume. This unique capability is essential for nonlinear coherent spectroscopic applications and ultrashort pulse metrology.              

\section{Experimental Setup}

The experiment utilizes a table-top high-harmonic source driven by Yb-fiber-laser technology and delivering short pulses with high photon flux in the XUV. Such sources can provide significantly higher repetition rates than the currently operating free-electron lasers based on superconducting linear accelerators \cite{Faatz2016} and offer very high temporal resolution due to the intrinsically small timing jitter \cite{Rothhardt2016}. Here we use a high harmonic source, which is conceptually similar to the one presented in \cite{Klas2021}, but operating at a reduced repetition rate of $\mathrm{100\,kHz}$. A $\mathrm{20\,W}$ average power Yb-fiber laser providing $\mathrm{250\,fs}$ pulses at 1030\,nm is frequency doubled in a $\mathrm{1\,mm}$ thick BBO crystal. The resulting $\mathrm{200\,fs}$, $\mathrm{515\,nm}$ pulses at $\mathrm{10\,W}$ average power are post compressed to sub-$\mathrm{20\,fs}$ using a nonlinear hollow-core fiber with a subsequent chirped-mirror compressor. High-harmonic generation (HHG) is driven by focusing these pulses to a spot size of $\mathrm{90\,\mu m}$ into an Ar gas jet resulting in a generated photon flux of $\approx\mathrm{2\cdot 10^{15}\,photons/s}$ in the $\mathrm{11^{th}}$ harmonic (H11) at $\mathrm{26.5\,eV}$ photon energy. Subsequently, separation of the driving laser and the generated XUV radiation is achieved using two grazing-incidence plates \cite{Pronin2011}, with an XUV reflectivity of $\mathrm{55\%}$ each, and additional $\mathrm{200\,nm}$ thick aluminum filters with a measured XUV transmission of $\mathrm{20\%}$ \cite{Hilbert2020}. Spatial and spectral characterization of the generated XUV radiation is performed by using the $\mathrm{90^{\circ}}$-reflection of a removable $\mathrm{1\,\mu m}$ aluminum foil by means of a flat-field grating spectrometer equipped with a CCD camera.\\

\begin{figure}[t!]
\centering
\includegraphics[scale=0.65]{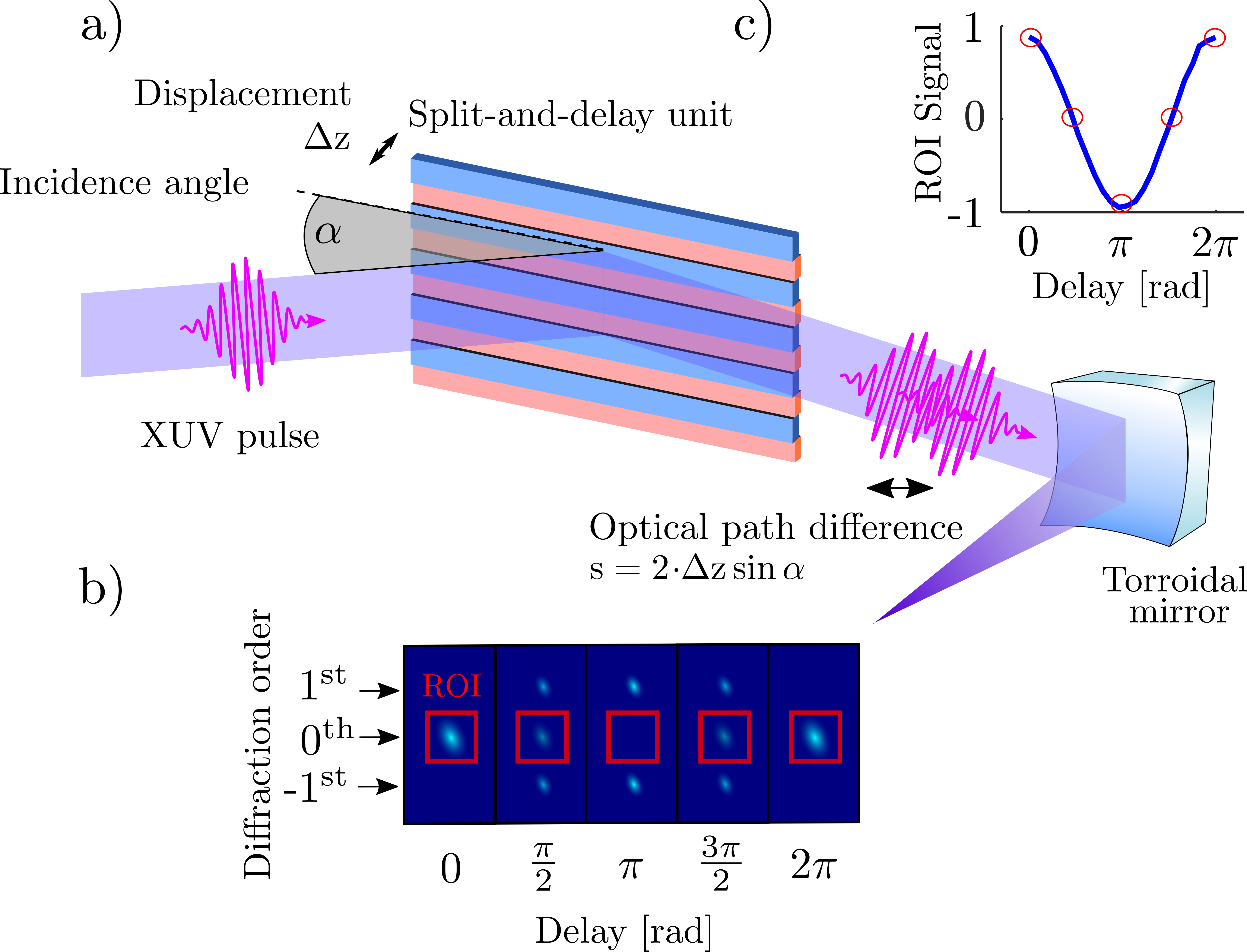}
\caption{Illustration of the experimental realization for XUV Fourier-transform spectroscopy. (a) An XUV pulse is reflected at $\mathrm{\alpha=8^{\circ}}$ incidence angle from a Michelson-type split-and-delay unit. The two coherent copies propagating collinearly are focused by a toroidal mirror onto a Ce:YAG scintillator. (b) In the focal plane at the Ce:YAG scintillator the diffraction orders of the lamellae can be resolved. Integrating the signal from the region-of-interest (red box, ROI) encompassing the $\mathrm{0^{th}}$ diffraction order as a function of relative phase delay yields the field-autocorrelation trace of the light source. (c) The Fourier transform of the sketched interferogram provides the spectrum of the XUV source.}
\label{fig:SDU}
\end{figure}

\newpage
Fig.\,\ref{fig:SDU} shows a schematic overview of the optical beam passing the FT instrument. The key element is the split-and-delay unit (SDU). It consists of two interleaved lamellar mirrors splitting the wavefront of the incoming XUV pulse uniformly across the beam profile. The design was suggested by Strong and Vanasse in the late 50's to overcome the absence of beam splitters in the far-infrared spectral range \cite{Strong1960}. In the 90's Möller proposed to extend the principle of lamellar mirrors to the high energy range\cite{Moeller1995}. The special mirror assembly in the present realization is cut from a silicon wafer and is coated with a $\mathrm{30\,nm}$ thick nickel layer. Its period is $\mathrm{d=250\,\mu m}$. The mechanical stability of the SDU is essential for interferometry at short wavelengths. In order to measure a high-quality interferogram of H11 at $\mathrm{\lambda=46.8\,nm}$ the relative position of the reflective elements should be known to a precision better than $\mathrm{\approx10\,nm}$, i.e. $\approx\lambda/4$. This is on the order of the background vibration in the laboratory, which is typically monitored during the measurements. The position of the reflective lamellar surface in the movable interferometer arm is controlled by 3-axes piezo-driven motors with nanometer precision. Building on previous demonstrations \cite{UsenkoASB17,UsenkoNAT17}, the setup is advanced byintegrating three commercial laser interferometers (SmarAct) enabling position measurement of the lamellar optics at repetition rates up to $\mathrm{40\,kHz}$ for online feedback control and offline FT data tagging. These refinements improve the precision of the relative lamellar-mirror displacement by 1-2 orders of magnitude, thereby improving the contrast in the recorded interferograms. This precision and stability is crucial for FT spectroscopy in the XUV spectral range. The XUV light is reflected from the autocorrelator mirrors at the grazing-incidence angle of $\mathrm{\alpha=8^{\circ}}$. The SDU provides collinear propagation of the two coherent pulse replicas with variable delay controlled by translating the movable lamellar mirror with nanometer and attosecond precision, respectively. Note that the optical path difference (OPD) is given by
\begin{equation}
\mathrm{s=2\cdot \Delta z\cdot \sin{\alpha}},
\end{equation}
with $\mathrm{\Delta z}$ being the mirror displacement. For instance, a displacement of $\mathrm{\Delta z = 1\,nm}$ corresponds to about $\mathrm{s=280\,pm}$, which results in a time delay between the copies of XUV pulses of only $\mathrm{\Delta t=0.93\,as}$.\\  

\begin{figure}[t]
\centering
\includegraphics[scale=0.43]{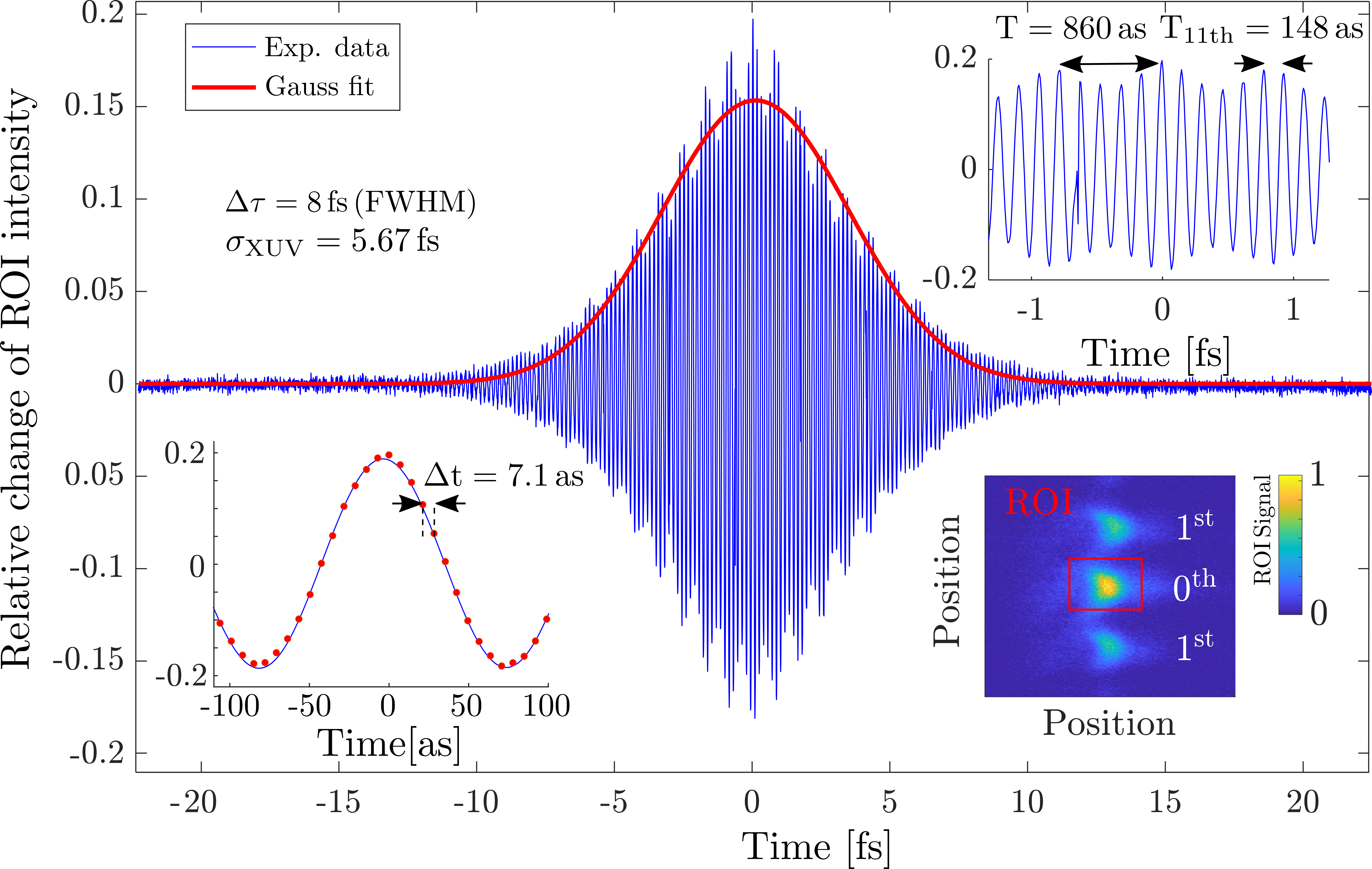}
\caption{Linear auto-correlation trace of the HHG source (blue line) measured with the Ce:YAG scintillator accumulating 50\,ms (5000 laser shots) per time step. A typical scintillator camera image of the diffraction orders is shown in the bottom right. The interferogram is derived by integrating the ROI (red square) and normalizing it to the intensity recorded by the entire CMOS detector. For sake of clarity we plot the difference between this signal and the average over the full delay scan. The Gaussian envelope fitted to the data gives a width $\mathrm{\Delta\tau=8\,fs}$ (FWHM) corresponding to a transform-limited pulse length of $\mathrm{\sigma_{XUV}=5.7\,fs}$. The magnifications show the field oscillations of the dominant $\mathrm{11^{th}}$ harmonic with a time period of $\mathrm{T_{11th}=148\,as}$ recorded with a sampling step size of $\mathrm{\Delta t=7.1\,as}$. In addition a slower beating pattern with a period of $\mathrm{T=860\,as}$ can be resolved.}
\label{fig:ACmeasurement}
\end{figure}

Upon reflection from the lamellar-mirror assembly, the XUV beam is focused onto a Ce:YAG scintillator by a toroidal mirror with a focal length of $\mathrm{f=5.3\,m}$. The wavefronts of each partial beam are parallel in each diffraction order. Thus, the temporal interference generated in the far-field depends only on the longitudinal displacement of the lamellar mirrors. The diffraction maxima are separated by $\mathrm{\Delta r = \lambda f/d}$, where $\mathrm{\lambda}$ is the central wavelength, $\mathrm{d}$ is the lamellar period and $\mathrm{f}$ is the focal length \cite{UsenkoASB17}. The maximum intensity continuously slips between even and odd diffraction orders as we scan the relative phase (time) delay between the two beams. At delay $\mathrm{\tau=0}$ the maximum intensity is reached in $\mathrm{0^{th}}$ order, whereas for $\mathrm{\tau=\pi}$ the odd diffraction orders are enhanced with maximum contrast due to the collinear beam propagation in the common-path interferometer \cite{UsenkoNAT17}. It allows for tight focusing of phase-coherent pulse replicas, while keeping the fringe contrast at maximum in the focal point. This unique capability is essential for nonlinear coherent spectroscopic applications striving to unravel ultrafast quantum phenomena in matter by detecting electrons and ions generated in the focal volume. For cameras and similar detectors this is not strictly true however, since it is in principle possible to select a small partial fraction of the fringe system. In the present experiment, the fluorescence signal from the scintillator is recorded by a CMOS camera. In order to obtain the field-autocorrelation (AC) trace of the light source, we integrate the signal intensity in the region-of-interest (ROI) encompassing the $\mathrm{0^{th}}$-order diffraction maximum. Note that the $\mathrm{0^{th}}$-order maximum consist of all frequencies selected by a plane multi-layer mirror, which was designed for a photon energy of $\mathrm{26.6\,eV}$ and installed behind the lamellar-mirror assembly in front of the vacuum chamber hosting the detectors. In data processing, we account for laser intensity fluctuations by normalizing the AC signal to the overall signal intensity recorded by the CMOS detector. Additionally, we apply a centroiding algorithm correcting the position of the image center in the far-field. This allows for correction of small beam-pointing fluctuations remaining from the beam-stabilization system of the optical driving laser. The statistical analysis of the recorded scintillator images gives a spatial jitter in the XUV focal plane of $\mathrm{\Delta x = 15\,\mu m}$ horizontally and $\mathrm{\Delta y = 60\,\mu m}$ (FWHM) vertically, i.e. in dispersion direction (see the inset in Fig.\,\ref{fig:ACmeasurement}). This is approximately a factor of 3 smaller than the beam waist at focus assuming diffraction-limited optics and Gaussian XUV beams. Beam-pointing instabilities on this order of magnitude are usually insignificant for applications using electron and ion spectrometers. The field autocorrelation $\mathrm{A_{F}(\tau)}$ is defined by

\begin{equation}
\begin{split}
    \mathrm{A_{F}(\tau)}&=\int\limits_{-\infty}^{+\infty}\vert E_{XUV}(t)+E_{XUV}(t-\tau)\vert^{2} \mathrm{dt}\\
    &=2\cdot \int\limits_{-\infty}^{+\infty}\vert E_{XUV}(t)\vert^{2} \mathrm{dt}+2\cdot Re \int\limits_{-\infty}^{+\infty}E^{\ast}_{XUV}(t)E_{XUV}(t-\tau)\mathrm{dt},
\end{split}
\label{eq:ACformula}
\end{equation}
where $E_{XUV}(t)$ represents the complex electric field of the XUV pulse and $\mathrm{\tau}$ is the delay between two laser pulses. Note that the first term refers to the pulse energy and the second is the field AC. Thus, Eq.(\,\ref{eq:ACformula}) is the foundation of Fourier-transform spectroscopy. Measuring the interferogram $\mathrm{A_{F}(\tau)}$ is equivalent to measuring the filtered spectrum of the harmonic source. The recorded signal is periodically modulated as a function of time delays and its Fourier transform yields the spectrum. We emphasize that the temporal profile (chirp) of the XUV pulse cannot be determined from $\mathrm{A_{F}(\tau)}$. However, the spectral bandwidth provides the transform-limited pulse duration and vice versa. For a drive pulse with a sufficiently small chirp, the interferogram can visualize the number of attosecond bursts in one attosecond pulse train emitted during the HHG process as described in \cite{Chen2014}.\\

The sampling energy width of the Fourier-transform spectrometer depends solely on the OPD between the two XUV pulse replicas
\begin{equation}
    \mathrm{\Delta E \propto 1/L},
    \label{eq:FTresolution}
\end{equation}
where L is the maximum optical path difference that can be set in the autocorrelator. In our instrument the corresponding maximum useful mirror displacement of $\mathrm{\Delta z=1\,mm}$ at a grazing-incidence angle of $\mathrm{\alpha=8^{\circ}}$ results in $\mathrm{L=278\,\mu m}$, which gives a maximum sampling energy width of $\mathrm{\Delta E = 4.5\,meV}$. We note that for the device depicted in Fig.\,\ref{fig:SDU} the given resolution is limited intrinsically by the transient loss of the transverse overlap between the two beams reflected from the lamellar-mirror assembly for larger delays, i.e. the increasing mirror displacement resulting in parallel beam separation.\\

Due to the linearity of the Fourier transformation, the absolute calibration of the instrument requires only one reference spectral line. However, if this anchor point is not available, precise calibration of the delay time axis, i.e. the optical path difference in both interferometer arms, is still possible. It mainly requires exact knowledge of the grazing-incidence angle of the photon beam impinging on the lamellar-mirror assembly. Under these experimental conditions, an error of $\mathrm{\Delta\alpha=0.1^{\circ}}$ would result in a relative energy mismatch of about $\mathrm{1.2\,\%}$, which is on the order of $\mathrm{300\,meV}$ for 26\,eV photon energy. Note that this relative error is uniform and constant overall photon energies. In contrast, grating spectrometers separate wavelengths according to their characteristic diffraction angles. Thus, its absolute calibration requires series of well-known emission lines distributed over a broad spectral range e.g. equally spaced harmonics from an HHG process. Although the calibration of FT seems to be straightforward, we note in passing that it requires a well-collimated beam to start with. In our experiment we use a single toroidal mirror as a collimator. It is known that a toroid produces coma aberration, which we resolve in the far-field image of the focal spot (Fig.\,\ref{fig:ACmeasurement}). Obviously, any kind of aberration in the far-field reduces the interference contrast in the measurement and a complex optical system for its compensation could be applied. For instance, studies using three toroidal mirrors controlled by a genetic algorithm achieved coma-free micrometer-size focal spots \cite{Frassetto2014}. However, it is the strength of the current common-path interferometer that these aberrations do not play a role here. As we will see in the following, as long as individual diffraction orders can be resolved in the far-field image the quality of the interferograms is extremely good.    

\section{Results}

The field AC trace of the HHG source is shown in Fig.\,\ref{fig:ACmeasurement}. It reveals periodic, sine-like modulations multiplied with a Gaussian envelope. The interferogram is obtained by integrating the intensity of the $\mathrm{0^{th}}$-order interference maximum and plotting the corresponding signal as a function of time delay between the two XUV pulses. A typical image recorded by the camera shows the $\mathrm{0^{th}}$ and $\mathrm{1^{st}}$ diffraction orders as displayed in the inset (on the bottom right). We fit a Gaussian envelope to the AC signal and obtain the autocorrelation width of $\mathrm{\Delta\tau = 8\,fs}$ (FWHM). Thus, the transform-limited XUV pulse length derived from the measurement is $\mathrm{\sigma_{XUV}=\Delta\tau/\sqrt{2} = 5.67\,fs}$. Note that this is in excellent agreement with simulations of the HHG phase matching window \cite{Constant1999} and the single-atom response \cite{Lewenstein1994}, resulting in an XUV pulse duration of $\mathrm{<6\,fs}$ \cite{Klas2021}.

\newpage
In order to demonstrate the stability of the experimental setup we recorded interferograms with a sampling step size of $\mathrm{\Delta t=7.1\,as}$. The clear oscillation shown in the zoom on the top right and bottom left in the Fig.\,\ref{fig:ACmeasurement} demonstrate the precise control of the relative phase of the harmonic pulses on the single-digit attosecond time scale. The field oscillation has a period $\mathrm{T_{11th}=148\,as}$. In addition, a superimposed beating pattern is visible in the interferogram with a longer period of $\mathrm{T=860\,as}$. It results from the neighboring harmonic orders H9 and H13, which pass the multi-layer mirror (ML) to some extent. It is worthwhile to note that this period matches the temporal separation of the individual bursts of attosecond laser pulses emitted at each optical half cycle of the driving laser pulse, whose visibility is suppressed due to the spectral filtering caused by the ML. At $\mathrm{515\,nm}$ drive wavelength, this corresponds to the observed time period of $\mathrm{860\,as}$.\\   

\begin{figure}[t!]
\centering
\includegraphics[scale=0.5]{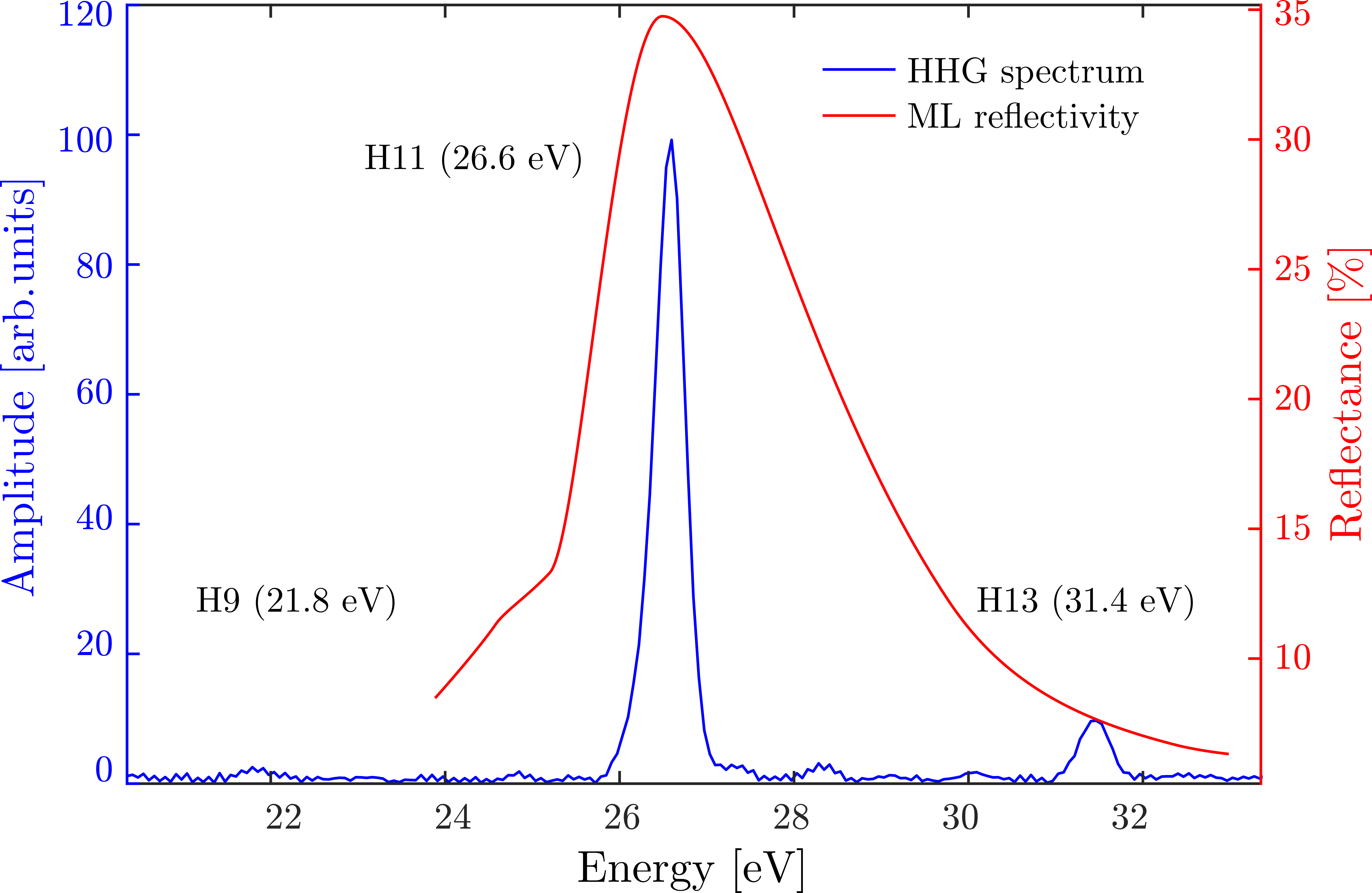}
\caption{The XUV spectrum of the HHG source obtained by the Fourier transform of the AC signal (blue). The spectrum reveals a dominant contribution from H11 at $\mathrm{26.6\,eV}$ and two satellite peaks from H9 ($\mathrm{21.8\,eV}$) and H13 ($\mathrm{31.4\,eV}$). The amplitude of satellites is strongly suppressed by the reflectance of the multi-layer mirror (red), which is designed for H11.}
\label{fig:HHGspectrum}
\end{figure}

The Fourier transform of the recorded interferogram is displayed in Fig.\,\ref{fig:HHGspectrum}. As expected, the spectrum shows a dominant peak at $\mathrm{26.6\,eV}$ (H11) with adjacent two weak satellite peaks at $\mathrm{21.8\,eV}$ (H9) and $\mathrm{31.4\,eV}$ (H13), respectively. This is in good agreement with the reflectivity curve of the multi-layer mirror, which was optimized for a photon energy of $\mathrm{26.6\,eV}$ selecting H11 and strongly suppressing the other frequencies from the harmonic comb.\\   

\begin{figure}[t!]
\centering
\includegraphics[scale=0.70]{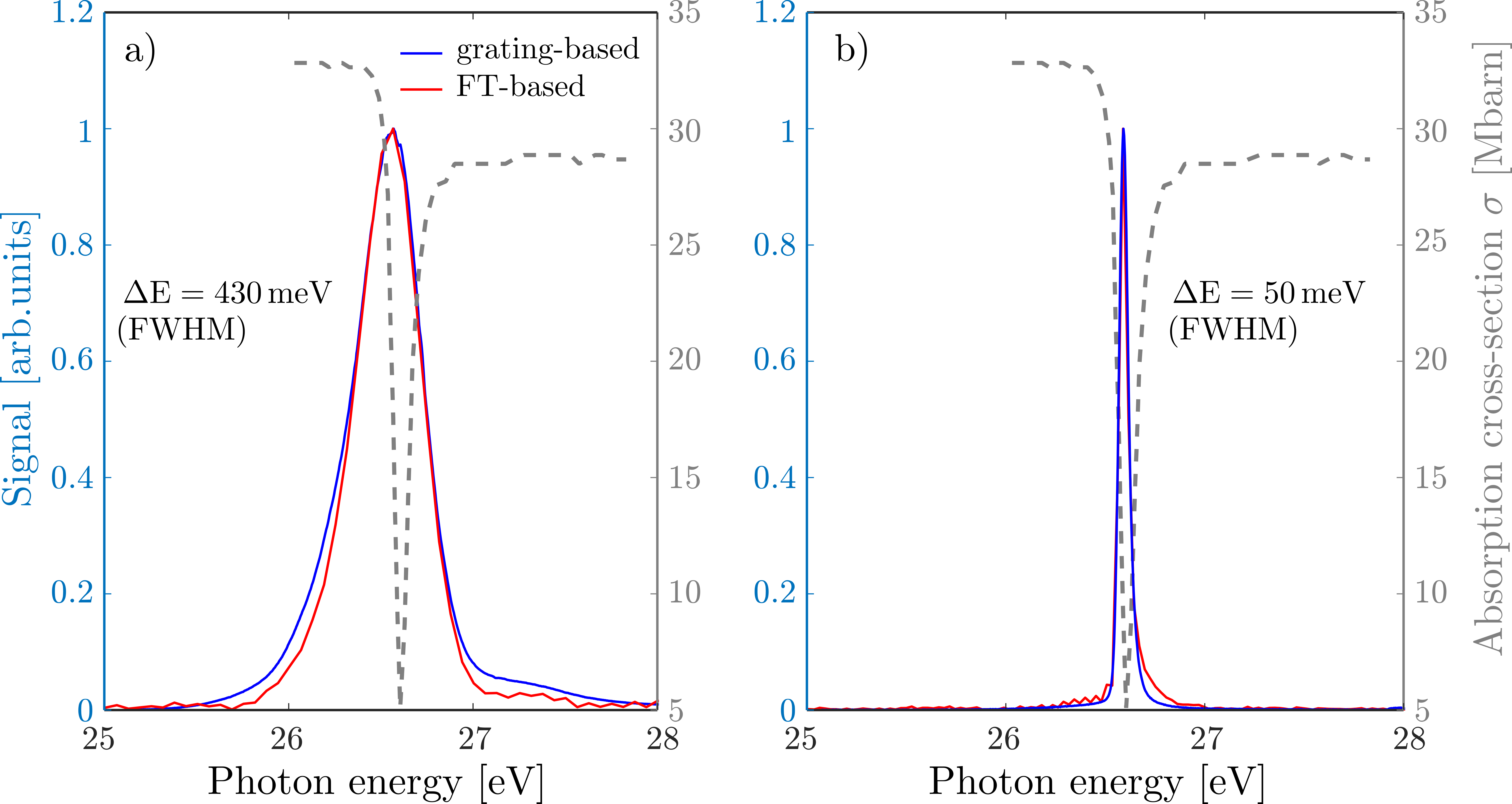}
\caption{Spectra of the H11 emission resolved with a high-resolution XUV spectrometer (blue) and retrieved from FT measurement (red), respectively. (a) Spectra measured at the base pressure $\mathrm{p=1\cdot 10^{-5}\,mbar}$. (b) Spectra measured at the Ar backing pressure of $\mathrm{p_{Ar}=8\cdot 10^{-2}\,mbar}$. The total absorption cross-section of Ar is plotted as a gray dashed line \cite{Madden1969}. The sharp minimum in the absorption is due to the $\mathrm{3s3p^{6}4p^{1}P^{1}}$ Fano resonance, which locally modifies the total cross-section \cite{Sorensen1994}.}
\label{fig:XUVEFTcomparison}
\end{figure}

To exploit the capacity of the FT instrument we spectrally narrow the $\mathrm{11^{th}}$ harmonic by filling an approximately 1\,m long part of the vacuum chamber with Ar to increase its partial pressure, i.e. atom density. The presence of the Fano resonance in the photoabsorption continuum of Ar around $\mathrm{26.6\,eV}$ modifies the transmitted spectrum due to the spectrally narrow decrease of the absorption cross-section \cite{Madden1969}. We have selected this transition for our benchmark experiment, because the underlying spectral tailoring has been carefully analyzed previously by means of high-resolution grating-based spectroscopy \cite{Rothhardt2014a}. In Fig.\,\ref{fig:XUVEFTcomparison} the results obtained from the FT measurements are compared with the spectrum recorded with the grating-based XUV spectrometer at two different conditions. The total absorption cross-section of Ar is plotted as a gray-dashed line. At a base pressure of (a) $\mathrm{p=1\cdot10^{-5}\,mbar}$, i.e. negligible absorption due to the residual gas, the H11 energy bandwidth is $\mathrm{\Delta E = 430\,meV}$ (FWHM). Its width is due to the short pulse duration (broad spectrum) of the $\mathrm{515\,nm}$ driving laser. Subsequently, we tailor the H11 transmission through the setup by utilizing the window-like Fano resonance. Therefore, we increase the Ar gas flow up to a backing pressure of (b) $\mathrm{p_{Ar}=8\cdot10^{-2}\,mbar}$. Under this condition both spectrometer types resolve the asymmetric Fano profile with $\mathrm{\Delta E = 50\,meV}$ (FWHM). The results are in very good agreement with the reported line width of $\mathrm{\Gamma=76\,meV}$ for the $\mathrm{3s3p^{6}4p^{1}P^{1}}$ Fano resonance measured at a synchrotron \cite{Sorensen1994}. Notably, the FT energy resolution is solely limited by the accessible delay range generated by the SDU as defined in Eq.(\,\ref{eq:FTresolution}). As described in section 2, the specifications of the FT instrument give an energy sampling width of $\mathrm{\Delta E = 4.5\,meV}$. This allows to achieve a resolution of about $\mathrm{\Delta E\approx 13\,meV}$ with three sampling points per spectral line at $\mathrm{26.6\,eV}$, which corresponds to $\mathrm{E/\Delta E}=2000$. Thus, the FT instrument offers comparable resolution as state-of-the-art high-resolution grating-based XUV spectrometers. The latter used in our experiment provides $\mathrm{\Delta E = 20\,meV}$ \cite{Hilbert2020} and its performance is in addition limited by the slit size, which compromises the signal strength on the detector.\\

An important parameter in Fourier-transform spectroscopy is the signal-to-noise ratio in the measured AC signal. To estimate the uncertainty limit for the position of spectral features we use the following empirical relation to estimate the line precision 
\begin{equation}
    \Delta \sigma\,[\mathrm{eV}]= \frac{k}{\sqrt{N_{W}}}\frac{W}{S/N}\cdot\frac{1}{\mathrm{8065\,cm^{-1}}},
\end{equation}
where $\mathrm{k}$ is a constant of the order of unity, which depends on the line shape. The value of $\mathrm{N_{W}}$ gives the number of statistically independent points in a fitted line width $\mathrm{W}$ (FWHM) and $\mathrm{S/N}$ is the signal-to-noise ratio \cite{Brault1987}. For the unperturbed H11 emission with $\mathrm{S/N=40}$, $\mathrm{N_{W}=7}$ and $\mathrm{W=430\,meV}$ the accuracy is about $\mathrm{4\,meV}$. For the sharp Fano resonance the accuracy is about $\mathrm{0.7\,meV}$. Obviously, this effect is significantly smaller than the current resolution limit set by the maximum beam path difference of one interferometer arm with respect to the other. 

\section{Summary}

In this contribution we present benchmark experiments applying a compact common-path interferometer for Fourier-transform spectroscopy in the extreme ultraviolet spectral range. The key component of the device is a broad-bandwidth Michelson-type all-reflective split-and-delay unit. It allows to record field-autocorrelation traces at XUV wavelengths as a function of time delay between two coherent pulse replicas with single-digit attosecond resolution, while scanning delays up to $\mathrm{1\,ps}$. The capacity of the technique is demonstrated by characterizing the $\mathrm{11^{th}}$ harmonic emission line of an ultrafast $\mathrm{515\,nm}$ driving laser. As we tailor the unperturbed $\mathrm{430\,meV}$ wide spectrum at $\mathrm{26.6\,eV}$ by the window-like $\mathrm{3s3p^{6}4p^{1}P^{1}}$ Fano resonance in Ar atoms we can clearly resolve the asymmetric lineshape with a natural width of $\mathrm{\Delta E = 50\,meV}$ (FWHM) by sampling the data with an energy sampling width of $\mathrm{4.5\,meV}$. We benchmark the results by comparing the data with experiments using a state-of-the-art grating-based high-resolution XUV spectrometer with a resolution of $\mathrm{20\,meV}$ and reach very good agreement. The power of interferometry for attosecond science becomes obvious when looking at the high-harmonic generation process in time domain. The recorded field-autocorrelation trace of the XUV light wave oscillation with a dominant time period of $\mathrm{148\,as}$ shows fingerprints of attosecond bursts emitted every half cycle of the optical drive pulse, i.e. every $\mathrm{860\,as}$.\\

\section{Outlook}

A future challenge for applications requiring simultaneously high laser power density at focus position and a spectral resolution exceeding a few times $10^{3}$ is to increase the effective surface of the lamellar-mirror assembly. This would support expanding the beam profile (larger numerical aperture) reflected from its surface and increasing the OPD without losing the spatial overlap and interference contrast, respectively. It calls for advanced engineering and materials design, because manufacturing lamellar mirrors with larger area without distortion of the substrate on a level of 10\,nm or less is demanding. Spatial distortions introduce phase errors in the reflected wavefronts, which reduce the interference contrast. Obviously, this is of great relevance for applications in the soft X-ray spectral range where the limits are truly on the single-digit nm scale. Therefore, we have started to develop special manufacturing protocols and nickel-coating procedures of the micro-machined silicon substrates for producing the next generation of lamellar-mirror assemblies for broadband soft x-ray applications. Recently, we have performed pioneering soft x-ray autocorrelation experiments, where we managed to control the relative phase of pulse replicas at 4.5\,nm central wavelength on the corresponding few-attosecond timescale \cite{Usenko2020}. We could show that soft x-ray interferometry is a powerful tool to probe ultrafast correlated quantum phenomena, such as the excitation of Auger shake-up states with sub-cycle precision. Even without recording a complete autocorrelation trace and just measuring a few optical cycles we discovered that specific non-radiative Auger-decay channels show a relative time (phase) delay of only $\mathrm{3\,as}$ with respect to the driving light wave oscillation. This is among the fastest electronic processes ever measured demonstrating the power of phase-sensitive detection of interaction products using coherent soft x-ray pulse pairs.\\

Thus we can conclude that the performance of our interferometric autocorrelator at the heart of the FT spectrometer opens up a new window of opportunities for soft X-ray pulse metrology, dynamic studies and nonlinear coherent spectroscopy in matter, materials and in the life sciences. The development of table-top setups based on coherent femtosecond and attosecond sources is fundamental in this challenging field, as the access to free-electron lasers or other large-scale facilities is limited for obvious reasons. This is especially true considering the recent progress offered by high-harmonic generation sources pushing spectroscopic studies in time and frequency domain to the ultimate limit given by the uncertainty principle connecting temporal and energy resolution. Of particular interest is the so-called water window, which refers to the spectral range between the K-absorption edge of oxygen at a wavelength of $\mathrm{2.34\,nm}$ and the K-absorption edge of carbon at $\mathrm{4.4\,nm}$ corresponding to photon energies of 530 and 280 eV, respectively. Water is transparent to these soft x-rays, while carbon, nitrogen and other elements found in biomolecules are absorbing, which may allow to unravel biological functions with element-specificity in natural environment.\\

\section*{Funding.}
This work was supported by the Fraunhofer Cluster of Excellence Advanced Photon Sources (CAPS), by the Innovation Pool of the Research Field Matter of the Helmholtz Association of German Research Centers in project (ECRAPS), by the Deutsche Forschungsgemeinschaft (DFG, German Research Foundation) through the Cluster of Excellence ‘Advanced Imaging of Matter’ (EXC 2056 - project ID 390715994), the collaborative research center ‘Light-induced Dynamics and Control of Correlated Quantum Systems’ (SFB-925 – project 170620586), the projects KI 482/20-1 and LA 1431/5-1 and by the Federal Ministry of Education and Research (BMBF) in project APPA R\&D: Licht-Materie Wechselwirkung mit hochgeladenen Ionen (13N12082).\\

\section*{Disclosures.}
The authors declare no conflicts of interest.\\

\section*{Data Availability.}
The authors declare that the main data supporting the findings of this study are available within the article. Extra data are available from the corresponding author upon reasonable request.\\



\bibliography{bibliography}






\end{document}